\documentclass[english,letterpaper,twocolumn,showpacs,pra,aps]{revtex4}
\usepackage{graphicx}
\usepackage{amssymb}

\makeatletter

\large  

\input epsf

\makeatother

\usepackage{babel}
\makeatother
\begin{document}

\title{Fermion space charge in narrow-band gap semiconductors, Weyl semimetals and around highly charged nuclei}

\author{Eugene B. Kolomeisky$^{1}$, Joseph P. Straley$^{2}$, and Hussain Zaidi$^{1}$}

\affiliation
{$^{1}$Department of Physics, University of Virginia, P. O. Box 400714,
Charlottesville, Virginia 22904-4714, USA\\
$^{2}$Department of Physics and Astronomy, University of Kentucky,
Lexington, Kentucky 40506-0055, USA}

\begin{abstract}
The field of charged impurities in narrow-band gap semiconductors and Weyl semimetals can create electron-hole pairs when the total charge $Ze$ of the impurity exceeds a value $Z_{c}e.$  The particles of one charge escape to infinity, leaving a screening space charge.  The result is that the observable dimensionless impurity charge $Q_{\infty}$ is less than $Z$ but greater than $Z_{c}$.  There is a corresponding effect for nuclei with $Z >Z_{c} \approx 170$, however in the condensed matter setting we find $Z_{c} \simeq 10$.  Thomas-Fermi theory indicates that $Q_{\infty} = 0$ for the Weyl semimetal, but we argue that this is a defect of the theory. For the case of a highly-charged recombination center in a narrow band-gap semiconductor (or of a supercharged nucleus),  the observable charge takes on a nearly universal value.  In Weyl semimetals the observable charge takes on the universal value $Q_{\infty} = Z_{c}$ set by the reciprocal of material's fine structure constant.
\end{abstract}


\maketitle

\section{Introduction}

The experimental success of quantum electrodynamics (QED) lies in the domain of small fields where observations impressively match the theoretical calculations based on perturbation theory in the fine structure constant $\alpha=e^{2}/\hbar c$.  In calculations involving bound states of a nucleus of charge $Ze$ the fine structure $\alpha$ constant additionally appears in the combination $Z\alpha$.  Even though $\alpha$ is small, $Z\alpha$ may not be, so that perturbative analysis can fail when $Z \alpha\gtrsim1$.  One of the most profound physical effects predicted to take place in this regime is the instability of the ground state (the vacuum) against creation of electron-positron pairs, resulting in a screening space charge of electrons with positrons leaving physical picture \cite{ZP}.  Experimental study of this effect has not been possible, as stable nuclei with $Z\gtrsim1/\alpha\approx 137$  have not been created, and attempts to look for positron production in a temporarily created overcritical system of slowly colliding Uranium nuclei have not been successful \cite{ZP}.

The goal of this paper is to demonstrate that in the condensed matter setting the corresponding problems are the impurity states in narrow-band gap semiconductors (NBGS) \cite{Keldysh} and Weyl semimetals (WS) \cite{AB}.  The advantage of these systems is that the charges and fields required to see the ground state instability are modest and readily achievable.  Some of the effects may have already been seen \cite{Keldysh} without fully appreciating what they represent.  The outline of this paper is as follows:  

In Section II the phenomenon of critical charge is first explained heuristically (IIA) followed by more precise semiclassical argument (IIB) that relates the effect to that of quantum-mechanical fall to the center \cite{LL3}.  Then the critical charge problem for the Coulomb potential modified at small distances is analyzed via dimensional analysis (IIC) which in Section III is employed to demonstrate the feasibility of observation of its condensed matter analog, instability with respect to creation of electron-hole pairs in semiconductors.  

In Section IV Thomas-Fermi (TF) theory of screening by space charge is derived and its deficiencies and a modification are discussed.  One of the by-products of the analysis is the conclusion that the observable charge of an arbitrary overcritical source in the WS case is always given by the reciprocal of the inverse fine structure constant for the material.  

In Sections V and VI the TF theory is solved in two steps - uniformly charged half-space $\rightarrow$ spherically symmetric charge distribution - so that the existence  of several physically different regimes of screening can be appreciated, and to establish a relationship with previous analysis \cite{numerical,MVP}.  The TF analysis leads to the conclusion that there is total screening in the WS case, through an argument that is parallel to Landau's "zero charge effect" in QED \cite{LL4}.   This would have readily observable consequences.  However, we will argue in section VII that this claim of complete screening is not right, so that there after all can be an observable charge.

\section{Critical charge in quantum electrodynamics}

The Dirac equation for an electron in vacuum in the field of a point charge $Ze$ (the Dirac-Kepler problem) becomes invalid for $Z>1/\alpha\approx137$ \cite{LL4}.  
 
\subsection{Heuristic Argument}

This feature can be heuristically understood by starting with classical expression for the energy of an electron of mass $m_{e}$ and momentum $p$ in the field of charge $Ze$
\begin{equation}
\label{Cl_energy}
\varepsilon = c \sqrt{p^{2}+m_{e}^{2}c^{2}}-\frac{Ze^{2}}{r}
\end{equation}
and trying to estimate the ground-state energy.  Since the electron position cannot be determined to better than $\hbar$ divided by the uncertainty of momentum, $p$ and $r\gtrsim \hbar/p$ entering Eq.(\ref{Cl_energy}) may be regarded as the typical momentum and size of the quantum state, respectively.  Then the state energy  can be estimated as
\begin{equation}
\label{energy_vs_p}
\varepsilon(p)\gtrsim  c \left (\sqrt{p^{2}+m_{e}^{2}c^{2}}-z p\right ), ~~z=Z\alpha
\end{equation}
where $z$ measures the nuclear charge in units of the reciprocal of the fine structure constant $1/\alpha$;  both these "natural" units (lower case letters) and the usual units for charge (upper case letters) will be used throughout this paper.  Minimizing with respect to the free parameter $p$ we find 
\begin{equation}
\label{gs_properties}
p_{0}\simeq \frac{m_{e}cz}{\sqrt{1-z^{2}}}, r_{0}\simeq\frac{\hbar}{p_{0}}\simeq \lambda \frac{\sqrt{1-z^{2}}}{z}, \lambda=\frac{\hbar}{m_{e}c}=\frac{r_{e}}{\alpha}
\end{equation}  
where $\lambda$ is the electron Compton wavelength that sets the scale for the uncertainty of measurement of the electron position and $r_{e}=e^{2}/m_{e}c^{2}$ is the classical electron radius.  The lowest (ground-state) energy is then
\begin{equation}
\label{gs_energy}
\varepsilon_{0}= m_{e}c^{2}\sqrt{1-z^{2}}
\end{equation}
While reproducing the well-known ground-state properties of a hydrogen-like atom in the non-relativistic $z\ll1$ limit  (\ref{gs_properties}), as well as (coincidentally) matching the exact expression for the ground-state energy (\ref{gs_energy}) based on the analysis of the Dirac equation \cite{LL4}, these arguments also predict that the minimum of (\ref{energy_vs_p}) only exists for $z<1$ ($Z<137$).  As $z\rightarrow 1-0$, the ground state becomes sharply localized ($r_{0}\rightarrow 0$), the typical electron momentum diverges ($p_{0}\rightarrow \infty$), and the ground-state energy vanishes ($\varepsilon_{0}\rightarrow 0$).  The conclusions (\ref{gs_properties})-(\ref{gs_energy}) become meaningless for $z>z_{c}=1$;  specifically the ground state energy is predicted to become imaginary.  The counterintuitive independence of $z_{c}$ of the electron mass $m_{e}$ can be explained via dimensional analysis:

The Dirac-Kepler problem is fully specified by the dimensionless parameter $z$, and by the electron Compton wavelength $\lambda$ in (\ref{gs_properties}).  If there exists a critical value of the charge $z_{c}$, it can only be a function of the remaining independent dimensionless parameters of the problem.  However having only one length scale $\lambda$ available makes it impossible to use it in a dimensionless combination.  Therefore $z_{c}$ cannot depend on $\lambda$, and thus on the electron mass $m_{e}$;  the only possible outcome is $z_{c}\simeq1$.  

These observations imply that the $z>1$ anomaly of the Dirac-Kepler problem persists in the Weyl-Kepler problem ($m_{e}=0$), where the estimate (\ref{energy_vs_p}) becomes      
\begin{equation}
\label{m=0_energy}
\varepsilon'(p)\simeq pc(1-z)
\end{equation}
As a result, a charged Weyl fermion placed in the field of a point charge with $z<1$ is always delocalized ($p_{0}=0$, $r_{0}=\infty$, and $\varepsilon_{0}=0$);  the spectrum is not discrete (i.e. no bound states).  On the other hand, a sufficiently attractive charge $z>1$ leads to a sharply localized ground state ($p_{0}=\infty$, $r_{0}=0$, and $\varepsilon_{0}=-\infty$).    

The $z>1$ instability in the Dirac-Kepler problem can be identified with a strong field limit of the Schwinger effect \cite{Schwinger}: the creation of electron-positron pairs in vacuum in a uniform electric field.  The phenomenon is characterized by the Schwinger typical electric field $E_{S}$ for which the work to separate the constituents of the electron-positron pair over the length scale of the Compton wavelength is equal to the rest energy of the pair: $eE_{S}\lambda\simeq m_{e}c^{2}$:  
\begin{equation}
\label{Schwinger}
E_{S}=\frac{m_{e}^{2}c^{3}}{e\hbar}
\end{equation}
For $E\lesssim E_{S}$ the pairs are created by tunneling with the vacuum being in a metastable state while for $E\gtrsim E_{S}$ the vacuum is absolutely unstable with respect to pair creation.  For the Coulomb problem the instability sets in when the electric field of the nucleus a Compton wavelength away from its center, $Ze/\lambda^{2}$, reaches the order of magnitude of the Schwinger field (\ref{Schwinger}), thus predicting $z_{c}\simeq1$.  In view of its mass independence, the prediction $z_{c}\simeq 1$ also applies to the Weyl-Kepler problem.  

\subsection{Critical charge as a consequence of quantum-mechanical "fall to the center"}

The $z >1$ anomaly of the Dirac equation is related to the "fall to the center" effect of quantum mechanics \cite{LL3}.

For a classical electron of energy $\mathcal{E}$ and angular momentum $M$ moving in a central field $U(r)$ the equation of conservation of energy can be written as
\begin{equation}
\label{nr_energy_conservation}
p_{r}^{2}=2m_{e}E - 2m_{e}U(r)-\frac{M^{2}}{r^{2}}>0
\end{equation}
where $p^{2}= p_{r}^{2}+M^{2}/r^{2}$ is the square of total momentum and $p_{r}$ is the radial component of the momentum.  The particle can reach the origin ("fall to the center") if  \cite{LL1}
\begin{equation}
\label{condition_to_fall}
\lim_{r\rightarrow 0}\left (r^{2}U(r)\right ) <-\frac{M^{2}}{2m_{e}}
\end{equation}
Specifically for zero angular momentum $M$ the fall to the center occurs for any attractive potential decreasing faster than $1/r^{2}$ as $r\rightarrow 0$.  In quantum mechanics $M^{2}$ has to be replaced by the eigenvalues of the square of the angular momentum operator which will be chosen in the semi-classical Langer form $M^{2}\rightarrow \hbar^{2} (l+1/2)^{2}$ \cite{ZP,LL3} so that both the effects of zero-point motion ($l=0$) and angular momentum ($l\neq 0$) are accounted for.   The smallest value of $M^{2}$ is $\hbar^{2}/4$ which implies that the fall to the center can occur for $1/r^{2}$ potentials that are more attractive than the critical potential satisfying \cite{LL3}
\begin{equation}
\label{cr_potential}
U_{c}(r\rightarrow 0)=-\frac{\hbar^{2}}{8m_{e}r^{2}}
\end{equation}
When this condition is met there is no lower bound on the spectrum.

For relativistic classical particle moving in a central field, energy and momentum are related by $\mathcal{E}=c\sqrt{p_{r}^{2} + M^{2}/r^{2}+ m_{e}^{2}c^{2}}+U(r)$.  For bound states we have 
$-m_{e}c^{2} <\mathcal{E} < m_{e}c^{2}$, where beyond the lower limit the system is unstable against pair creation.  At the lower limit the range of motion can be found by solving for the radial momentum to obtain a form analogous to Eq.(\ref{nr_energy_conservation}):
\begin{equation}
\label{r_energy_conservation}
p_{r}^{2}=\frac{1}{c^{2}}\left (U^{2}(r)+2m_{e}c^{2}U(r)\right )-\frac{M^{2}}{r^{2}}>0
\end{equation}
If $U(r)$ is diverging as $r\rightarrow 0$, the $U^{2}(r)$ term dominates, and for attractive $U(r)$ the origin can be reached if \cite{LL2}
\begin{equation}
\label{r_confition_to_fall}
\lim_{r\rightarrow 0}\left (rU(r)\right )<-Mc
\end{equation}
Classically the fall to the center for the $M=0$ state is possible for a potential that at $r\rightarrow 0$ is more attractive than a $1/r$ potential.  The quantum case is different:  substituting in (\ref{r_confition_to_fall}) minimal $M^{2}=\hbar^{2/}4$ we infer that the fall to the center is possible for $1/r$ potentials more attractive than the critical potential
\begin{equation}
\label{rel_cr_potential}
U_{c}(r\rightarrow 0)=- \frac{\hbar c}{2r}
\end{equation}
Comparing with the Coulomb field $U=-Ze^{2}/r$, we deduce $z_{c}=Z_{c}\alpha=1/2$ which is the correct critical charge of the Kepler problem for a spinless particle \cite{ZP}.  It is half the value found for the Dirac particle (the "missing" half is due to the electron spin).  The Dirac case cannot be fully understood semi-classically but an insight can be gained by observing that the relativistic case with $\mathcal{E}=-m_{e}c^{2}$ is equivalent to a non-relativistic problem with the effective potential (compare Eqs.(\ref{nr_energy_conservation}) and (\ref{r_energy_conservation}))
\begin{equation}
\label{eff_potential}
U_{eff}(r)=-\frac{U^{2}(r)}{2m_{e}c^{2}}-U(r) + \frac{M^{2}}{2m_{e}r^{2}} 
\end{equation}   
and zero total energy.  We now see that for the Kepler problem,  $U(r)=-Ze^{2}/r$, the particle is \textit{repelled} at large distances.  The same effective potential is obtained from the Klein-Gordon equation transformed into the Schr\"odinger form \cite{ZP}.  

The Dirac equation can be also brought into a Schr\"odinger form with an effective potential at $\mathcal{E}=-m_{e}c^{2}$ resembling Eq.(\ref{eff_potential}) but also exhibiting extra terms attributed to the electron spin \cite{ZP}.   Their role  can be (approximately) summarized in a form similar to Eq.(\ref{eff_potential}) with different amplitude of the $1/r^{2}$ term.  Specifically for the Dirac-Kepler problem we have \cite{ZP}
\begin{equation}
\label{eff_potential_dirac}
U_{eff}(r)=\frac{Ze^{2}}{r}+\frac{\hbar^{2}(1-z^{2})}{2m_{e}r^{2}}
\end{equation}
We see that the fall to the center occurs for $z >1$ and then the particle is confined to the central region of radius
\begin{equation}
\label{localization_scale}
R_{cl}= \frac{\lambda(z^{2}-1)}{2z}
\end{equation}

As is the case of the fall to the center problem, proper treatment of the instability of the Dirac equation requires accounting for finite radius $a$ of atomic nucleus that modifies the $1/r$ attraction at short distances and removes the difficulty for $z>1$ \cite{ZP}.  For a nucleus with $Z=Z_{c}\approx170$ ($z_{c}\approx 1.24$) the ground-state energy reaches the boundary of the lower continuum, $\varepsilon_{0}=-m_{e}c^{2}$ \cite{ZP}.  Past that point the total energy of the production of an electron-positron pair becomes negative and the vacuum becomes unstable with respect to pair creation;  the positron repelled by the nucleus escapes to infinity while the electron remains near the nucleus \cite{ZP}.  

\subsection{Dimensional analysis}

Many of the conclusions of previous analysis of the critical charge problem that accounted for finite radius $a$ of atomic nucleus \cite{ZP,Popov76} can be reproduced by a combination of dimensional analysis and simple physical arguments.   Indeed now we have a problem fully specified by independent dimensionless combinations $Z$, $\alpha$, and $a/\lambda$.  Then if there exists a critical value $Z_{c}$, it can only be a function of $\alpha$ and $a/\lambda$.  The electrostatic potential inside the nucleus has the form:  $\varphi(r)= (Ze/a)G(r/a)$ where $G(1)=1$ and $G(0)$ is finite.  Then the parameters $Z$ and $\alpha$ appear together in the $z=Z\alpha$ combination.  Therefore
\begin{equation}
\label{dim_analysis_zeta}
z_{c}=f\left (\frac{a}{\lambda}\right )
\end{equation}
or 
\begin{equation}
\label{dim_analysis}
Z_{c}=\frac{1}{\alpha}f\left (\frac{a}{\lambda}\right )=\frac{\hbar c}{e^{2}}f\left (\frac{m_{e}ca}{\hbar} \right )
\end{equation}
where $f(y)$ is a function that depends on the shape of the charge distribution within the nucleus.  The properties of $f(y)$ can be inferred from the following arguments: 

(i) For $a=0$ one has $z_{c}=1$ which implies the small argument behavior $f(y\rightarrow 0)\rightarrow 1$.  This additionally means that $z_{c}=1$ for $m_{e}=0$  for arbitrary $a$ (the Weyl-Kepler problem with cutoff).  

(ii) In the classical $\hbar\rightarrow 0$ limit the Planck's constant must drop out of Eq.(\ref{dim_analysis}).
This translates into the large argument behavior $f(y\rightarrow \infty)\simeq y$ with the consequence $z_{c}\simeq a/\lambda$.  This is indeed what is expected on physical grounds:  the vacuum becomes unstable when the electron potential energy at the center of the nucleus $-e\varphi(0) +m_{e}c^{2}$ reaches the boundary of the lower energy continuum $-m_{e}c^{2}$.  This argument applied to the uniformly charged ball model of the nucleus predicts $f(y\rightarrow \infty)\rightarrow 4y/3$.  On the other hand, $f(y\rightarrow \infty)\rightarrow 2y$ for the constant potential ball model of the nucleus.

For ordinary heavy nuclei the nuclear size $a$ depends on $Z$ according to the Fermi formula 
\begin{equation}
\label{nucl_matter}
a=0.61 r_{e}Z^{1/3}=0.61\lambda \alpha^{2/3}z^{1/3}
\end{equation}
The critical charge $z_{c}$ is found by simultaneous solution of Eqs.(\ref{dim_analysis_zeta}) and (\ref{nucl_matter}) for $z=z_{c}$: 
\begin{equation}
\label{Fermi_model_electron}
z_{c}=f\left (0.61\alpha^{2/3}z_{c}^{1/3}\right )
\end{equation} 

The electron Compton wavelength is known to be much larger than the nuclear size which means the argument of the function $f$ in Eq.(\ref{Fermi_model_electron}) is much smaller than unity.  In this limit the distinction between different models of nuclear charge distribution disappears thus explaining the nearly model-independent value of $z_{c}$ close to $1$.

To add credibility to dimensional analysis we now show that the latter easily solves the problem of instability of the muon vacuum.  Indeed, the muon Compton wavelength has the same order of magnitude as the nuclear size since the muon is more than $200$ times heavier than the electron.  The solution to the problem is then described by Eq.(\ref{dim_analysis}) with the electron mass $m_{e}$ replaced by the muon mass $m_{\mu}$.  We are now in the large argument limit, $f(y\rightarrow \infty)\simeq y$, which determines the critical $Z$ for the muon to be
\begin{equation}
\label{muon_charge}
Z_{c}^{(\mu)}=\frac{z_{c}^{(\mu)}}{\alpha}\simeq \left (\frac{m_{\mu}}{m_{e}}\right )^{3/2}\simeq 3000
\end{equation}

These conclusions agree with existing analysis of the problem \cite{ZP, Popov76}.  In fact we observe that approximating the true $f(y)$ dependence for \textit{all} $y$ by its $y\gg1$ limit \cite{Popov76}
\begin{equation}
\label{choice_of_f}
f(y)=\frac{4}{3}y+ 1.1547, ~~~~~~~y\gg1
\end{equation}
suffices to quickly estimate the critical charge in the practically relevant case of the uniform density model of the nucleus.  In the $y\ll1$ limit, where this approximation is expected to work poorly, the combination of Eqs.(\ref{Fermi_model_electron}) and (\ref{choice_of_f}) predicts $Z_{c}=163$ which is close to the accepted value of $170$.  Inspection of previous results \cite{ZP,Popov76} shows that, except for a narrow vicinity of $y=0$, the function $f(y)$ is basically a straight line of the right slope with a larger-than-unity offset.  The significance of the linear approximation (\ref{choice_of_f}) is that it allows us to reliably estimate the critical charge in the condensed matter setting. 

\section{Critical charge in condensed matter setting}

QED's predictions of screening by space charge can be tested in performable experiments involving condensed matter systems, both presently available and those that will become available in the near future.  Our primary example is that of NBGS whose physics is known to mimic QED \cite{ZP}.  Excitation of an electron-hole pair is analogous to the creation of an electron-positron pair in QED, with the band gap representing the combined rest energy of the particles.  Creation of the electron-hole pairs in the presence of a uniform electric field takes place via Zener tunneling \cite{Zener} which is analogous to the Schwinger effect \cite{Schwinger}.  Our contention is that moderately charged impurity regions in semiconductors can trigger a space charge around them that parallels the effects that would occur in QED for unrealistically large $Z\gtrsim170$.  

The idea that the $Z>137$ anomaly of the original Dirac-Kepler problem may have observable condensed matter implications is due to Keldysh \cite{Keldysh}.  In his study of the impurity states in semiconductors Keldysh noted that the effective mass approximation \cite{LL9}, while successful in describing shallow impurity states,  fails to explain deep states whose binding energy is comparable with the band gap.  Such states are formed near multi-charged impurity centers, vacancies etc. and they cannot be associated with either conduction or valence bands.  The experiment presented another puzzle:   some highly charged impurities acted as very efficient recombination centers that managed to trap both electrons and holes but an explanation why that was the case was lacking.  Keldysh argued that experimental findings can be explained in a two-band approximation (well obeyed in NBGS of the $InSb$ type) where the low-energy electron (hole) dispersion law is relativistic \cite{Keldysh,Kane}
\begin{equation}
\label{dispersion}
\varepsilon(\textbf{p})=\pm \sqrt{(\Delta/2)^{2}+v^2 p^{2}}.
\end{equation}            
Here the upper and lower signs correspond to the conduction and valence bands, respectively, $\Delta\equiv 2mv^{2}$ is the energy band gap that parallels twice the rest energy of a particle of mass $m$, and $v$ is the velocity of a high-momentum particle analogous to the speed of light $c$.  Compared to their vacuum electron-positron counterparts, electrons and holes in NBGS have two orders of magnitude smaller effective mass ($m\simeq 0.01m_{e}$) and limiting velocity $v$ nearly three orders of magnitude smaller than the speed of light ($v\approx 4.3 \times 10^{-3}c$).  As a result their band gap $\Delta\simeq 0.1 eV$ is seven orders of magnitude smaller than the rest energy of the electron-positron pair \cite{Zawadzki}.  Due to these parameter values the analog of large field QED effects are readily realizable in NBGS.

With this in mind, the determination of the impurity states reduces to solving the Dirac equation for a particle of mass $m$ in the field of a charge $Ze$ screened by the dielectric constant $\epsilon$ of the semiconductor.  In view of the peculiarity of the Dirac-Kepler problem (now $\alpha=e^{2}/\hbar v\epsilon$), Keldysh argued that for $z=Z\alpha < 1$ the impurity states are given by the known solution to the Dirac equation \cite{LL4} while the $z>1$ case with "collapsed" ground state describes a recombination center.  

An expression for the NBGS critical charge can be written in a form that parallels Eq.(\ref{dim_analysis_zeta}):
\begin{equation}
\label{sc_dim_analysis}
z_{c}=f\left (\frac{a}{\Lambda}\right ),~~~~ \Lambda=\frac{\hbar}{mv}=\frac{2\hbar v}{\Delta}=\frac{R_{e}}{\alpha}
\end{equation}
where now $a$ is the radius of the impurity region, $\Lambda$ is the semiconductor analog of the electron Compton wavelength and $R_{e}=2e^{2}/\epsilon\Delta$ is the semiconductor counterpart of the classical electron radius (defined as band electron's delocalization size at which its potential self-energy $e^{2}/\epsilon R_{e}$ matches its rest energy $mv^{2}=\Delta/2$.  We note that since both $R_{e}\simeq1 nm$ and $\Lambda\simeq 10 nm$ significantly exceed the lattice spacing, macroscopic theory of impurity states ignoring the lattice structure of the material suffices.  

The density of nuclear matter is known to have the order of magnitude set by the classical electron radius $r_{e}$ (see Eq.(\ref{nucl_matter})).  It will be made clear shortly, that the large field effects in NBGS become prominent at impurity charge densities set by the NBGS electron radius $R_{e}$.  Therefore the relationship between the radius $a$ and the charge $Z$ of a uniformly charged region will be chosen as \begin{equation}
\label{sc_matter}
a=1.3 R_{e} Z^{1/3}=1.3 \Lambda \alpha^{2/3}z^{1/3}
\end{equation}
that parallels its nuclear physics counterpart (\ref{nucl_matter});  the numerical factor corresponds to the charge density $n_{ext}=10^{20}cm^{-3}$ to be justified below.  The NBGS critical charge can be determined by solving the equation
\begin{equation}
\label{sc_unif_den_model}
z_{c}=f\left (1.3\alpha^{2/3}z_{c}^{1/3}\right ), 
\end{equation}
that is nearly identical to its QED counterpart (\ref{Fermi_model_electron}).  Because the value of the limiting velocity $v$ is known, the semiconductor equivalent of the fine structure constant is $\alpha=e^{2}/\epsilon\hbar v  \approx 1.7/\epsilon= 0.17$, an order of magnitude larger than its QED counterpart  (we employed $\epsilon=10$ \cite{Kittel}).  With this value of $\alpha$ and choosing the function $f(y)$ in the simple form (\ref{choice_of_f}), the solution to Eq.(\ref{sc_unif_den_model}) is $z_{c}\approx1.7$ which implies $Z_{c}\approx10$.  The corresponding critical cluster size is $a_{c}\approx3nm$ according to Eq.(\ref{sc_matter}) .  Surely $Z\gtrsim 10$ impurity clusters with sizes in excess of several nanometers are more common objects than $Z\gtrsim170$ nuclei.  
 
In addition to making it possible to study the regime of large effective fine structure constant, condensed matter systems also offer possibilities that cannot be realized in QED.  Indeed, over forty years ago Abrikosov and Beneslavski\u{i} \cite{AB} predicted the existence of WS having points in the Brillouin zone where the valence and conduction bands meet with a dispersion law that is linear in the momentum.  This is the $\Delta =0$ case of Eq.(\ref{dispersion}). The low-energy excitations in WS (realizing massless versions of QED) are described by the Weyl equation.  We already know that the critical charge for the Weyl-Kepler problem is $z_{c}=1$.   Eq.(\ref{m=0_energy}) additionally implies lack of the discrete spectrum for $z < 1$;   for $z>1$, a space charge of Weyl fermions is present in the ground state.  Such substances are likely to be realized in doped silver chalcogenides $Ag_{2+\delta}Se$ and $Ag_{2+\delta}Te$ \cite{silver}, pyrochlore iridates $\mathcal{A}_{2}Ir_{2}O_{7}$ (where $\mathcal{A}$ is Yttrium or a lanthanide) \cite{pyro}, and in  topological insulator multilayer structures \cite{topo}.  The zero energy gap of WS implies that in a \textit{uniform} electric field the creation of a space charge of Weyl fermions is spontaneous.  The dielectric constant of WS is of order $10$ with $e^{2}/\hbar v\simeq1 $ \cite{AB}, thus leading, like in the NBGS case, to $Z_{c}\simeq10$ independent of the size of the impurity region.

NBGS and WS are condensed matter systems where analogs of the atomic collapse of QED can be experimentally detected.  Related phenomena can be also observed in graphene.  Indeed, graphene possesses the linear dispersion law analogous to that of WS and microscopic parameters similar to NBGS which leads to a small value for the critical charge for promotion of electrons from the valence band to the conduction band.  Such a problem has been considered elsewhere \cite{graphene} and experimental signatures of the "atomic collapse" in graphene were recently reported \cite{collapse}.  The graphene problem is mathematically different from what we discuss, because graphene is a two-dimensional semimetal embedded in a three-dimensional space.    

Below we will determine the ground-state properties of NBGS and WS in the presence of a finite-size positive Coulomb impurity (a negative charge leads to the same discussion due to particle-hole symmetry). The arguments given above imply that at modest $Z$ electrons are promoted from the valence band to form a space charge around the impurity while the holes leave the physical picture; the properties of the space charge vary with $Z$ and $\alpha$ and are determined by the interplay of attraction to the impurity (promoting the creation of electron-hole pairs), and electron-electron repulsion combined with the Pauli principle (limiting the creation of the space charge).  The QED analysis of the physical properties of the space charge was carried out in two limits:

(i) $Z$ close to $Z_{c}$, where there are very few electrons promoted to the conduction band for which the single-particle picture is a good starting point \cite{ZP}; and

(ii) $Z\gg Z_{c}$, where the number of screening electrons is large and the electron-electron interactions cannot be ignored \cite{numerical,MVP}.

Below we demonstrate that the physics in the $Z\gg Z_{c}$ limit exhibits a large degree of universality.  Although we are mostly concerned with the NBGS setting, our findings are equally applicable in QED as both problems share the same mathematics;  a solution to the WS problem benefits the understanding of the NBGS/QED case.   

To help the readers orient themselves between three physically different manifestations of the problem and to provide them with a condensed matter-QED translation dictionary, in Table I we summarized pertinent properties of electrons in vacua of QED, NBGS and WS.  
\begin{table}[htdp]
\begin{center}
\begin{tabular}{llll}\textit{Media} & \textit{QED} &  \textit{NBGS} & \textit{WS} \\ Electrons & free & band Dirac & band Weyl \\ Mass & $m_{e}$ & $m\simeq10^{-2}m_{e}$ & $0$ \\Degeneracy $g$ & $2$ &$\geqslant2$ & $\geqslant 2$\\Dielectric & $\epsilon=1$& $\epsilon \approx 10$&$\epsilon\simeq10$\\constant \\ Limiting speed & $c$ & $v\approx 4 \times 10^{-3}c$ & $v\simeq10^{-2}c$\\ Band gap or &$2m_{e}c^{2}$ &$10^{-7}\times2m_{e}c^{2}$&$0$\\rest energy\\Fine structure & $\frac{e^{2}}{\hbar c}\approx\frac{1}{137}$ &$\frac{e^{2}}{\hbar v\epsilon}\approx \frac{1}{6}$&$\frac{e^{2}}{\hbar v\epsilon}\simeq 0.1$\\constant $\alpha$\\Coupling &$\frac{4\alpha^{3}}{3\pi}\approx10^{-7}$&$\frac{2g\alpha^{3}}{3\pi}\lesssim10^{-3}$&$\frac{2g\alpha^{3}}{3\pi}\lesssim10^{-3}$\\constant $\gamma$\\Classical radius&$r_{e} \simeq10^{-6}nm$ & $R_{e}\simeq 1nm$&$\infty$\\of electron\\Compton&$\lambda\simeq10^{-4}nm$&$\Lambda\simeq10nm$&$\infty$\\wavelength\\ Schwinger or&$E_{S}\simeq10^{16}\frac{V}{cm}$&$E_{Z}\simeq10^{5}\frac{V}{cm}$&0\\Zener field
\end{tabular} 
\caption{Summary of properties of electrons in vacua of quantum electrodynamics (QED), narrow band gap semiconductors (NBGS) and Weyl semimetals (WS).}
\end{center}
\label{dictionary}
\end{table}
The entries not yet specified are:  

(i)   \textit{The fermion degeneracy factor $g$} which is $2$ in QED while in NBGS it is twice the number of Dirac valleys (\ref{dispersion}) within the first Brillouin zone; an isotropic valley-independent limiting velocity $v$ is assumed for simplicity.   In the WS case $g$ counts the number of Weyl points within the first Brillouin zone:  $g=24$ in pyrochlore iridates \cite{pyro} and  $g=2$  in a topological insulator multilayer \cite{topo}.  

(ii) \textit{The coupling constant $\gamma$} plays a role analogous to that of the fine structure constant $\alpha$ in  polarization effects, as will be made clear below. 

(iii) \textit{The Zener field $E_{Z}$} is the semiconductor analog of the Schwinger field (\ref{Schwinger}) defined as
\begin{equation}
\label{Zener}
E_{Z}=\frac{\Delta^{2}}{e\hbar v}
\end{equation}
Comparing the values of the fields $E_{S}$ and $E_{Z}$ explains why NBGS are so well suited to study strong field QED effects; the situation is even more favorable in WS where due to the zero band gap, an arbitrarily weak field is "strong" as far as the space charge phenomena are concerned.

\section{Thomas-Fermi theory}

Since for $Z\gg Z_{c}$ a large number of electrons are in the conduction band, the properties of the system consisting of the impurity and its interacting cloud of electrons can be understood semi-classically with the help of the TF theory \cite{numerical,MVP}.  The main object of the latter is a physical electrostatic potential $\varphi(\textbf{r})$ felt by an electron that is due to both the electrostatic potential of the impurity $\varphi_{ext}(\textbf{r})$ and that of the space charge characterized by the number density $n(\textbf{r})$:
\begin{equation}
\label{definition_potential}
\varphi(\textbf{r})=\varphi_{ext}(\textbf{r})-\frac{e}{\epsilon}\int \frac{n(\textbf{r}')dV'}{|\textbf{r}-\textbf{r}'|}
\end{equation}
The external potential $\varphi_{ext}(\textbf{r})$ is a pseudopotential that represents the perturbation of the system caused by the impurity; even though $\varphi_{ext}$ is not entirely of electrostatic origin, we will define $\triangle \varphi_{ext} = -4\pi e n_{ext}/\epsilon$.  We assume that the impurity charge density $en_{ext}(\textbf{r})$ is spherically-symmetric and localized within a mesoscopic region of size $a$ so that for $r \geqslant a$ the potential $\varphi_{ext}(\textbf{r})$ reduces to a purely Coulomb form $\varphi_{ext}(r)= Ze/\epsilon r$ of a net charge $Ze$.  There are several reasons why the impurity region has to be mesoscopic in size:

First of all, in practice charged atomic scale defects cannot have $Z\gtrsim10$.    Second, a large charge localized within a small region implies a large electrostatic potential.  However all our analysis is based on approximating the exact dispersion law by its low-energy limit (\ref{dispersion}).  For that to remain valid the order of magnitude of the potential within the impurity region should not exceed a volt.  Like in graphene, this corresponds to the electronvolt energy scale which is significantly smaller than the width of the conduction band.  Finally the conditions of semi-classical description inherent within the TF theory must be met.  All these constraints along with the requirement  $Z\gg Z_{c}\simeq10$ can be satisfied in $a\gtrsim 10 nm$ impurity clusters.  Promotion of electrons to the conduction band also takes place in smaller (down to $3nm$) regions but the number of these electrons may not be large enough for the predictions of the TF theory to be quantitatively reliable.      

Given $\varphi(\textbf{r})$, one can deduce that the electron number density $n(\textbf{r})$ is different from zero only in the region of space where the electron potential energy $-e\varphi(\textbf{r}) +\Delta/2$ drops below $-\Delta/2$, thus defining the edge of the space charge region as
\begin{equation}
\label{shell_region}
e\varphi(\textbf{r})> \Delta,~~~n(\textbf{r})>0
\end{equation}    
The radius of the space charge region $R_{sc}\geqslant a$ is given by the equalities $e\varphi(R_{sc})=\Delta$, $n(R_{sc})=0$;  outside the region we have $n=0$ and 
\begin{equation}
\label{observable_charge_definition}
\varphi=\frac{Q_{\infty}e}{\epsilon r},~r>R_{sc}=\frac{1}{2}Q_{\infty}\frac{2e^{2}}{\epsilon \Delta}\equiv \frac{1}{2}Q_{\infty}R_{e}
\end{equation}
where $Q_{\infty}<Z$ is the observable charge  as seen at large distances from the source center.  Continuity of the potential $\varphi$ across the shell boundary relates $R_{sc}$ and $Q_{\infty}$ while the NBGS electron radius $R_{e}$ sets the length scale as indicated in the last two steps in (\ref{observable_charge_definition}) meaning that we can speak of the shell size or the observable charge interchangeably.  We note that in the WS case, $R_{e}=\infty$,  Eq.(\ref{observable_charge_definition}) predicts $R_{sc}=\infty$, i.e. the electron shell extends all the way to infinity.  In natural units of charge the relationship between the observable charge $q_{\infty}=Q_{\infty}\alpha$ and the radius of the electron shell is given by
\begin{equation}
\label{qed_observable_charge_definition}
q_{\infty}=\frac{2R_{sc}}{\Lambda}
\end{equation}

From the thermodynamical standpoint, creation of electron ($e$)-hole ($h$) pairs by the field of a Coulomb impurity accompanied by escape of a hole to infinity may be viewed as a "chemical reaction" $e+h\leftrightarrows 0$ (the ground state of the semiconductor is the "vacuum") \cite{LL9};  the condition of equilibrium for this reaction has the form
\begin{equation}
\label{chem_potentials}
\mu_{e}+\mu_{h}=0,\mu_{e}=\sqrt{(\Delta/2)^{2}+v^2 p_{F}^{2}}-e\varphi,\mu_{h}= \Delta/2
\end{equation}
where $\mu_{e}$ and $\mu_{h}$ are the chemical potentials of the electrons and holes, respectively, and
\begin{equation}
\label{Fermi_momentum}
p_{F}(\textbf{r})=\hbar \left (\frac{6\pi^{2}n(\textbf{r})}{g}\right )^{1/3}
\end{equation}
is the Fermi momentum which we assume is a slowly varying function of position $\textbf{r}$.  

The condition of equilibrium (\ref{chem_potentials}) together with Eq.(\ref{Fermi_momentum}) implies a relationship between the physical potential and the number density of the space charge \cite{numerical,MVP}:
\begin{equation}
\label{n_of_phi}
n(\textbf{r})= \frac{\gamma}{4\pi} \bigg\{\frac{\epsilon \varphi(\textbf{r})}{e}\frac{\epsilon}{e^{2}}[e\varphi(\textbf{r})-\Delta]\bigg\}^{3/2}
\end{equation} 
where 
\begin{equation}
\label{gamma}
\gamma = \frac{2g\alpha^{3}}{3\pi}
\end{equation}
is the coupling constant  characterizing the relative strength of electron-electron interactions and zero-point motion.  The four orders of magnitude disparity between its condensed matter and QED values (see Table I) is yet another indication that the space charge phenomenon is more relevant to semiconductors than to QED.   

Since the electron chemical potential in (\ref{chem_potentials}) is set at the boundary of the lower continuum, in the NBGS/ QED cases the screening of the external charge is incomplete;  only in the WS ($\Delta=0$) case do we have complete screening.  The latter statement can be rigorously proven by setting $\Delta=0$ in (\ref{chem_potentials}) and combining the outcome with Eqs.(\ref{definition_potential}) and (\ref{Fermi_momentum}):
\begin{equation}
\label{Weyl_int_eq}
\left (\frac{4\pi n(\textbf{r})}{\gamma}\right )^{1/3}-\frac{\epsilon \varphi_{ext}(\textbf{r})}{e}+\int\frac{n(\textbf{r}')dV'}{|\textbf{r}-\textbf{r}'|}=0
\end{equation}    
Taking in Eq.(\ref{Weyl_int_eq}) the $r\rightarrow \infty$ limit gives a relationship
\begin{equation}
\label{condition_on_n}
\int n(\textbf{r})dV=Z\Big\{1-\lim_{r\rightarrow \infty}\left (\frac{4\pi n(\textbf{r})r^{3}}{\gamma Z^{3}}\right )^{\frac{1}{3}}\Big\}
\end{equation}
whose consequences are that the electron number density $n(\textbf{r})$ must decay faster than $r^{-3}$ at $r$ large and that
\begin{equation}
\label{n_normalization}
\int n(\textbf{r})dV=Z
\end{equation} 
i.e. according to TF theory, a WS succeeds in giving complete screening of the impurity charge.   

Applying the Laplacian operator to both sides of Eq.(\ref{definition_potential}) and using (\ref{n_of_phi})  we find the relativistic TF equation 
\begin{equation}
\label{diff_eq_phi}
\nabla^{2} \left (\frac{\epsilon \varphi}{e}\right ) =-4\pi n_{ext} +\gamma \bigg\{\frac{\epsilon \varphi}{e} \frac{\epsilon}{e^{2}}(e\varphi-\Delta)\bigg\}^{3/2}
\end{equation}
that was investigated in QED for the case of localized source term \cite{numerical,MVP}.  

Applicability of the zero-temperature TF theory to experiments to be conducted at finite temperature requires further justification.  Since the $0.1eV$ energy gap of NBGS significantly exceeds the room temperature scale of $1/40 eV$, the zero temperature theory adequately describe room temperature experiments.  However WS have a zero energy gap and in equilibrium the conduction and valence bands are populated with electrons and holes, respectively.  This effect can be also neglected as we are working at potential close to $1V$ significantly exceeding the room-temperature energy scale.

\subsection{Range of applicability of the Thomas-Fermi theory and proposal for its improvement}

In order to establish the range of applicability of the TF theory we note that the observable charge $q_{\infty}$ is the critical charge of the single-particle problem for a charged region of scale $R_{sc}$ which is due to both the external charge and that of the space charge.  Then replacing $a\rightarrow R_{sc}$ in Eq.(\ref{sc_dim_analysis}) we arrive at the definition
\begin{equation}
\label{exact_definition_obs_charge}
q_{\infty}=f\left (\frac{R_{sc}}{\Lambda}\right )
\end{equation}
that is consistent with the TF result (\ref{qed_observable_charge_definition}) only in the classical limit $R_{sc}\gg\Lambda$ (recall that $f(y\rightarrow\infty)\simeq y$, Section II).  However WS are characterized by $\Lambda=\infty$ and so the semiclassical condition can never be met.  The consequence is that the prediction of complete screening in the WS case, $q_{\infty}=Q_{\infty}\alpha=0$, an exact consequence of the TF theory, is an artifact.  In reality the WS will screen an overcritical impurity charge by space charge only until the point at which the single-particle description is restored.  This sets a limit on the applicability of the TF theory to the WS case at large distances from the source center and implies that the space charge region has a finite radius to be estimated below.  

Since the Weyl-Kepler problem with $z<1$ does not have a discrete spectrum, Coulomb impurities in WS can never be fully neutralized.  Their observable charge $q_{\infty}$ is either $1$ (and then there is a space charge of Weyl electrons) or $z<1$ (when there is no space charge).     

More generally Eq.(\ref{exact_definition_obs_charge}) implies that in the NBGS/QED setting the screening is never so great (in the case $z > z_{c}$) that the observable charge $q_{\infty}$ is less than unity.  In the point charge limit $a\rightarrow 0$ the space charge region must also shrink to a point  ($R_{sc}\rightarrow 0$), and then Eq.(\ref{exact_definition_obs_charge}) predicts $q_{\infty}=1$ i.e. disallowance for a point charge to have observable charge exceeding unity \cite{numerical}.  

Since the Dirac-Kepler problem always has bound states, Coulomb impurities in NBGS can be neutral or ionized with the outer electron shells partially or fully filled with ($Z> Z_{c}$) or without ($Z<Z_{c}$) the space charge being present.  Here we only consider the problem of an overcritical $Z> Z_{c}$ ion with all outer shells empty.     

To summarize, the condition of applicability of the TF theory can be stated in two equivalent forms, $R_{sc}\gg \Lambda$ or $q_{\infty}\gg1$.  Ultimately, the TF treatment of the space charge in the presence of the supercritical source $z\gg z_{c}$ is applicable because the fine structure constant is significantly smaller than unity.

In order to see what physics is missing from the TF theory, we observe that the relationship between the observable charge and the radius of the electron shell (\ref{qed_observable_charge_definition}) resembles the $z\gg1$ limit of the semi-classical expression (\ref{localization_scale}) for the localization scale of an electron (replacing $\Lambda\rightarrow \lambda$ and $q_{\infty}\rightarrow z$).  The latter is sensitive to the fall to the center occurring at $z=1$ while the TF result (\ref{qed_observable_charge_definition}) is not.      The same can be seen more generally by solving the main equation of the TF theory (\ref{chem_potentials}) relative to $p_{F}^{2}$:
\begin{equation}
\label{solution for p_F}
p_{F}^{2}=\frac{1}{v^{2}}\left ((e\varphi)^{2}-e\varphi \Delta \right )
\end{equation}
With the identifications $-e\varphi\rightarrow U$, $v\rightarrow c$, $\Delta\rightarrow 2m_{e}c^{2}$, this is the energy relationship Eq.(\ref{r_energy_conservation}), but the term involving the angular momentum is missing.  We propose including this into the right hand side of Eq.(\ref{diff_eq_phi}): 
\begin{eqnarray}
\label{modified_TF}
\frac{1}{r^{2}}\frac{d}{dr}\left (r^{2}\frac{d}{dr}\frac{\epsilon \varphi}{e}\right )&=&-4\pi n_{ext}(r)\nonumber\\
&+& \gamma \bigg\{\frac{\epsilon \varphi}{e} \frac{\epsilon}{e^{2}}(e\varphi-\Delta)-\frac{1}{\alpha^{2}r^{2}}\bigg\}^{3/2}
\end{eqnarray} 
At the boundary of the space charge region where the potential is given by Eq.(\ref{observable_charge_definition}) the expression in the second line vanishes leading to a relationship between observable charge $q_{\infty}$ and the radius of the space charge region $R_{sc}$
\begin{equation}
\label{charge_vs_radius_beyond_TF}
q_{\infty}=\frac{R_{sc}}{\Lambda}+ \sqrt{\left (\frac{R_{sc}}{\Lambda}\right )^{2}+1}
\end{equation}   
generalizing the TF result (\ref{qed_observable_charge_definition}) and  correctly capturing the limiting cases of $R_{sc}\gg\Lambda$ and $R_{sc}\ll \Lambda$.  This is equivalent to choosing the function $f$ in (\ref{exact_definition_obs_charge}) in the $f(y)=y+\sqrt{y^{2}+1}$ form.  Assessing the status of our modification of the TF theory (\ref{modified_TF}) requires a separate investigation which we postpone until the future.  At the very least what is proposed qualifies as an interpolation.  Until we learn more about the deficiencies of the TF theory, we focus on its standard version accumulated in Eqs.(\ref{n_of_phi}) and (\ref{diff_eq_phi}).  Some other desirable features of the modified TF equation (\ref{modified_TF}) are mentioned in Section VII.

\section{Strong screening regime:  $\gamma Z^{2}\gg1$}

The phenomenon of screening is a manifestation of the electron-electron interactions quantified by the  coupling constant $\gamma$. Its smallness (see Table I) does not imply that the screening response is necessarily weak.  Previous analysis \cite{MVP} established that the strength of screening is determined by the dimensionless combination $\gamma Z^{2}$ which for $Z\gg Z_{c}$ can take on an arbitrary value.  Below we additionally show that in the NBGS/QED cases the regime of strong screening further subdivides into that of super-strong screening $Z\gg \gamma^{-3/2}$ (that is hardly accessible in practice) and the regime of moderately strong screening $\gamma^{-1/2}\ll Z \ll \gamma^{-3/2}$ which is treated in detail.   

\subsection{Uniformly charged half-space}

Since the source region is mesoscopic  in size we find it useful to start with the problem of screening of a uniformly charged half-space ("half-infinite" nuclear matter in the QED setting) by space charge.  This requires solution of the one-dimensional version of Eq.(\ref{diff_eq_phi}):
\begin{equation}
\label{1d_diff_eq}
\frac{d^{2}}{dx^{2}} \left (\frac{\epsilon \varphi}{e}\right ) =-4\pi n_{ext}(x) +\gamma \bigg\{\frac{\epsilon \varphi}{e} \frac{\epsilon}{e^{2}}(e\varphi-\Delta)\bigg\}^{3/2}
\end{equation}
where $n_{ext}(x<0)=3Z/4\pi a^{3}$ while $n_{ext}(x>0)=0$.  The solution to Eq.(\ref{1d_diff_eq}) within the source region far away from the boundary $\varphi(x\rightarrow -\infty)\equiv\varphi_{-\infty}$ corresponds to the state of local neutrality $n=n_{ext}$ \cite{numerical}:
\begin{equation}
\label{phi_-infty}
\frac{\epsilon\varphi_{-\infty}}{e}R_{e}=1+\sqrt{1+\left (\frac{4\pi n_{ext}R_{e}^{3}}{\gamma}\right )^{2/3}}
\end{equation}
The quantity $e\varphi_{-\infty}$ is the work function, i.e. the energy needed to remove the electron from the source region.  At this point we observe that in the QED ($R_{e}\rightarrow r_{e}$, $n_{ext}r_{e}^{3}\simeq1$, $\gamma\ll1$) and WS ($R_{e}=\infty$) versions of the problem,  Eq.(\ref{phi_-infty}) simplifies to    
\begin{equation}
\label{phi_-infty_m=0}
\varphi_{-\infty}=\frac{e}{\epsilon}\left (\frac{4\pi n_{ext}}{\gamma}\right )^{1/3}
\end{equation}
In the condensed matter setting our theory is applicable provided the potential $\varphi_{-\infty}$ does not exceed a volt.  Then for $\gamma\simeq10^{-3}$ and $\epsilon\simeq10$ the maximal external charge density within the impurity region can be estimated as $n_{ext}\simeq10^{20} cm^{-3}$, two orders of magnitude smaller than the free-electron density in normal metals.  This corresponds to $n_{ext}R_{e}^{3}\approx 0.1$ and justifies Eq.(\ref{sc_matter}).  With these parameter values the approximation (\ref{phi_-infty_m=0}) also holds in NBGS.  Given the charge concentration of $n_{ext}=10^{20}cm^{-3}$, a $10 nm$ impurity region would contain a bare charge of about $400$ which is significantly larger than $Z_{c}\approx10$.  Yet larger values of $Z\gg Z_{c}$ can be obtained by choosing $a\gtrsim10nm$:  a $20 nm$ region will host an external charge of about $3,000$.  

The source boundary represents a perturbation to the constant $n_{ext}$;  assuming the effect is weak  we substitute $\varphi=\varphi_{-\infty}(1-\phi)$, $0\leqslant \phi\ll1$, into Eq.(\ref{diff_eq_phi}) and linearize about $\varphi=\varphi_{-\infty}$:  
\begin{equation}
\label{1d_linearized_TF}
\frac{d^{2}\phi}{dx^{2}}-\kappa^{2}\phi=0
\end{equation}
where the length scale
\begin{eqnarray}
\label{TF_length}
 \kappa^{-1}&=&3^{-1/2}(4\pi n_{ext})^{-1/3}\gamma^{-1/6}=\frac{e}{\sqrt{3\gamma}\varphi_{-\infty}\epsilon}\nonumber\\
 &=&3^{-5/6}(\gamma Z^{2})^{-1/6}a\simeq R_{e} \gamma^{-1/6}
 \end{eqnarray}
parallels the Debye length of the TF theory of screening in a Fermi gas \cite{Kittel};  $\kappa^{-1}$ is the scale over which the potential $\varphi$ recovers to $\varphi_{-\infty }$ when disturbed by an inhomogeneity.  The last estimate in (\ref{TF_length}) is only applicable to the NBGS or QED ($R_{e}\rightarrow r_{e}$) cases.  In the condensed matter setting with $\gamma \simeq10^{-3}$ the TF  screening length is of the order several nanometers.  In QED $\kappa^{-1}$ is an order of magnitude larger than the classical electron radius.

Applicability of the concept of the screening length to a finite size system is limited by the constraint $\kappa^{-1}\ll a$ which is a statement of strong screening $\gamma Z^{2}\gg1$ \cite{MVP}.  The crossover in the screening response occurs at a charge
\begin{equation}
\label{weak_strong_xover_charge}
Z_{x}\simeq \gamma^{-1/2}
\end{equation}
In condensed matter applications we find $Z_{x}\simeq 30$ -- both the regimes of weak $10\lesssim Z\lesssim 30$ and strong $Z\gtrsim 30$ screening are experimentally accessible.  In QED we have $Z_{x}\simeq 3000$ which is only of academic interest.  Assuming the potential at the impurity boundary is not significantly smaller than $\varphi_{-\infty}$, the $\varphi(x<0)$ dependence can be inferred from the linearized form (\ref{1d_linearized_TF}), which gives $\varphi_{-\infty}-\varphi(x<0)\propto e^{\kappa x}$:  while local neutrality holds far away from the boundary, it is violated in a boundary layer whose size has the order of magnitude of the TF screening length $\kappa^{-1}$.           

Outside the impurity region $x>0$ the potential will continue to decrease from its value at the boundary $\varphi(x=0)$ until it reaches the edge of the space charge region defined as $e\varphi(x=L_{sc})=\Delta$.  The length scale $L_{sc}$ has a meaning of the thickness of the layer of space charge outside the impurity region.  

We thus see that a layer of net positive charge of thickness $\kappa^{-1}$ localized next to the boundary is followed by a layer of negative charge of thickness $L_{sc}$ outside the source region \cite{numerical,MVP}. The net charge of this double layer is positive thus implying that the electric field for $x>L_{sc}$ is finite and uniform.  

Even though Eq.(\ref{1d_diff_eq}) can be integrated in quadratures,  an approximate solution is more illuminating.  Within the source region the potential is approximated by the solution to Eq.(\ref{1d_linearized_TF}) finite at $x=-\infty$:
\begin{equation}
\label{phi_x<0}
\frac{\epsilon \varphi}{e}=\frac{\epsilon \varphi_{-\infty}}{e}(1-\phi)=\frac{\kappa}{\sqrt{3\gamma}}(1- Ae^{\kappa x})
\end{equation}          
where it is assumed (and later justified) that $A\ll1$.

Outside the impurity region $x>0$ the full non-linear equation (\ref{1d_diff_eq}) becomes
\begin{equation}
\label{full_TF_x>0}
\frac{d^{2}}{dx^{2}} \left (\frac{\epsilon \varphi}{e}\right ) =\gamma \bigg\{\frac{\epsilon \varphi}{e} \frac{\epsilon}{e^{2}}(e\varphi-\Delta)\bigg\}^{3/2}
\end{equation}

In the WS case or when $e \varphi \gg \Delta$, Eq.(\ref{full_TF_x>0}) simplifies to 
\begin{equation}
\label{full_TF_x>0_Weyl}
\label{diff_eq_phi_m=0}
\frac{d^{2}}{dx^{2}}\left (\frac{\epsilon \varphi}{e}\right ) =\gamma \left ( \frac{\epsilon \varphi}{e}\right )^{3}
\end{equation}
An analytic solution to the problem of screening of a supercharged nucleus in the strong screening limit $\gamma Z^{2}\gg1$ that approximates the finite nucleus by half-infinite nuclear matter and relies on Eqs.(\ref{1d_linearized_TF})  and (\ref{full_TF_x>0_Weyl}) was proposed by Migdal, Voskresenski\u{i}, and Popov (MVP) \cite{MVP}.  

For $x>0$ the solution to Eq.(\ref{full_TF_x>0_Weyl}) satisfying the conditions of zero electric field and zero potential at $x=\infty$ has the form 
\begin{equation}
\label{phi_x>0}
\frac{\epsilon \varphi}{e}=\sqrt{\frac{2}{\gamma}}\frac{1}{x+B}
\end{equation}
Continuity of the potential and of the electric field at the source boundary $x=0$ determines the integration constants $A$ and $B$ in Eqs.(\ref{phi_x<0}) and (\ref{phi_x>0}) to be
\begin{equation}
\label{int_constants1}
A\approx 0.2374, ~~B=\beta\kappa^{-1}\simeq (\gamma Z^{2})^{-1/6}a, ~~\beta \approx 3.212
\end{equation}
The length scale $B$ naturally has the order of magnitude of the TF screening length $\kappa^{-1}$.

The profile of the electron number density for $x>0$ is implied by Eqs.(\ref{n_of_phi}) and (\ref{phi_x>0}):
\begin{equation}
\label{1d_density}
n(x)=\frac{1}{2\pi}\sqrt{\frac{2}{\gamma}}\frac{1}{(x+B)^{3}}
\end{equation}
In the $x\gg B\simeq \kappa^{-1}$ limit the MVP solution (\ref{phi_x>0}) and (\ref{1d_density}) exhibits universality, i.e. it becomes independent of the parameters of the source region.    

\subsection{Spherically-symmetric charge distribution}

The MVP solution Eqs.(\ref{phi_x<0})-(\ref{1d_density}) with $x\rightarrow r-a$ is partly relevant to the problem of screening response of NBGS or WS to the spherically symmetric charge distribution of radius $a$.  Specifically, Eq.(\ref{phi_x<0}) adequately solves the problem within the impurity region in the strong-screening regime $\gamma Z^{2}\gg1$.  For example, the net charge within the source region can be estimated as \cite{MVP}
\begin{equation}
\label{Q(r<a)}
Q(r\leqslant a)\simeq \kappa^{-1}a^{2}(Z/a^{3})\simeq Z(\gamma Z^{2})^{-1/6}
\end{equation}
This is significantly smaller than the bare charge $Z$ thus illustrating substantial screening of the source region.

On the other hand, the density profile Eq.(\ref{1d_density}) with $x\rightarrow r-a$ integrates to an infinite charge in three dimensions.  Therefore outside a spherically symmetric  charge distribution Eqs.(\ref{phi_x>0}) and (\ref{1d_density}) are only applicable as long as approximating the spherical surface by a plane is valid, i. e. for $x=r-a\ll a$.  

In order to go beyond the limitation of the MVP approximation outside the source region we need to solve the full three-dimensional equation (\ref{diff_eq_phi}). 

\subsubsection{Weyl semimetal}

Outside of the source in the WS ($\Delta=0$) case one has to look at the full non-linear equation (\ref{diff_eq_phi}) whose radially-symmetric solution is sought in the form 
\begin{equation}
\label{def_chi}
\frac{\epsilon \varphi(r)}{e}= \frac{1}{r}\chi \left (\frac{r}{a}\right )  
\end{equation}
where, via Gauss's theorem, the function $\chi$ is related to the charge $Q(r)$ within a sphere of radius $r$ as:
\begin{equation}
\label{charge_connection}
Q(r)=-r^{2}\frac{\partial(\epsilon\varphi/e)}{\partial r}=\chi(\ell)-\chi'(\ell),~~~\ell=\ln\frac{r}{a}
\end{equation} 
Substituting (\ref{def_chi}) into (\ref{diff_eq_phi}) for $r>a$ and $\Delta=0$ we obtain the equation
\begin{equation}
\label{diff_eq_chi_of_x}
\chi''(\ell)-\chi'(\ell)=\gamma \chi^{3} .
\end{equation}
For $\ell=\ln(r/a)\ll1$ we can neglect here the first-order derivative term $\chi'(\ell)$ compared to $\chi''(\ell)$;  then $Q(r)\approx -\chi'(\ell)$.  The solution to (\ref{diff_eq_chi_of_x}) in this limit is the MVP result (\ref{phi_x>0}) in disguise
\begin{equation}
\label{chi_near_boundary}
\chi_{1}(\ell)=\sqrt{\frac{2}{\gamma}}\frac{1}{\ell+B/a}, ~~~0\leqslant \ell\ll 1
\end{equation}    
In the strong-screening limit $\gamma Z^{2}\gg1$ the parameter $B/a\simeq (\gamma Z^{2})^{-1/6}$ drops out of Eq.(\ref{chi_near_boundary}), and the solution to the full Eq.(\ref{diff_eq_chi_of_x}) has the form $\chi(\lambda, \ell)=(2/\gamma)^{1/2}y(\ell)$ where $y(\ell)$ is a parameter free universal function satisfying singular boundary condition $y(\ell\rightarrow 0)\rightarrow \ell^{-1}$.  The latter behavior is no longer an accurate representation of the true dependence $y(\ell)$ past $\ell\simeq1$.  Therefore the solution (\ref{chi_near_boundary}) is only applicable up to a crossover scale $\ell=\ell^{*}\simeq1$, i.e. within several source radii as was already observed earlier.  Within this range the rescaled potential $\epsilon \varphi/e$ drops from a value of the order $\gamma^{-1/2}(\gamma Z^{2})^{1/6} a^{-1}$ at the source boundary to $\gamma^{-1/2}a^{-1}$ at the crossover scale $\ell^{*}$, and the charge within a sphere of radius $r=ae^{\ell}$ drops from the value given by Eq.(\ref{Q(r<a)}) at the source boundary to $Q^{*}\simeq -\chi_{1}'(1)\simeq \gamma^{-1/2}$ at the crossover scale $\ell^{*}$.

For $\ell=\ln(r/a)\gg1$ we can neglect in Eq.(\ref{diff_eq_chi_of_x}) the second-order derivative term $\chi''(\ell)$ compared to $\chi'(\ell)$;  then $q(\ell)=Q(r)\alpha\approx \chi(\ell)\alpha$ and  for arbitrary screening strength and in natural units of charge Eq.(\ref{diff_eq_chi_of_x}) acquires a form
\begin{equation}
\label{GL}
\frac{dq}{d\ell}=-\frac{2g\alpha}{3\pi} q^{3}
\end{equation}
that is mathematically identical to the Gell-Mann-Low (renormalization-group) equation \cite{LL4} for the physical charge in QED reflecting the effects of vacuum polarization.  Eq.(\ref{GL}) exhibits the Landau "zero charge" effect \cite{LL4}:  for any "initial" value of $q$ the system "flows" to the zero charge fixed point $q=0$ as $\ell\rightarrow \infty$ ($r\rightarrow \infty$), i.e. the source charge has been completely screened.  Alternatively,  for $r$ fixed complete screening is reached in the point source limit $a\rightarrow 0$.  Zero observable charge is an exact property of the TF theory discussed in Section IV.  

In the strong-screening regime the last equation is applicable at $\ell \gtrsim \ell^{*}\simeq 1$.  As a result the charge $q(r)$ inside a sphere of radius $r>a^{*}=ae^{\ell^{*}}\gtrsim a$ will be given by 
\begin{equation}
\label{0_charge_solution}
q^{2}(r)=\frac{z^{*2}}{1+(4g\alpha z^{*2}/3\pi)\ln(r/a^{*})}\rightarrow \frac{3\pi}{4g\alpha  \ln(r/a^{*})}
\end{equation}
where the integration constant $z^{*}=Q^{*}\alpha$ is the charge within a sphere of radius $a^{*}$.  Since $\gamma Q^{*2} \simeq1$, the constant can be estimated as $z^{*}\simeq\alpha^{-1/2}\gg1$ thus implying that the observable charge will be accurately given by the last representation in (\ref{0_charge_solution}) at distances exceeding several impurity radii.  Substituting $Q=q/\alpha=\chi$ into Eqs.(\ref{def_chi}) and (\ref{n_of_phi}) we find corresponding expressions for the potential
\begin{equation}
\label{W_potential_strong_screening}
\varphi(r)\approx \frac{e}{\epsilon r\sqrt{2\gamma \ln(r/a^{*})}}=\frac{e}{2\epsilon r}\sqrt{\frac{3\pi}{g\alpha^{3}\ln(r/a^{*})}}
\end{equation}
and the electron density
\begin{equation}
\label{W_density_strong_screening}
n(r)\approx \frac{\gamma}{4\pi r^{3}}\frac{1}{[2\gamma\ln(r/a^{*})]^{3/2}}=\frac{1}{16\pi r^{3}}\sqrt{\frac{3\pi}{g\alpha^{3}}}\ln^{-3/2}\left (\frac{r}{a^{*}}\right )
\end{equation}
both valid for $r\gtrsim a^{*}\simeq a$.  We note that due to the logarithmic factor the density profile is integrable.  This feature, a reflection of the three-dimensional character of the problem, is missing from the MVP result (\ref{1d_density}).

The hallmark of Eqs.(\ref{0_charge_solution})-(\ref{W_density_strong_screening}) is their near universality: a weak logarithmic dependence on the source size $a\simeq a^{*}$ with universal amplitudes.  Since both $a$ and $a^{*}$ appear within arguments of the logarithm, in what follows for simplification purposes the difference between them will be neglected.  We conclude that for $\gamma Z^{2}\gg1$ the solution to the screening problem within several impurity radii from the source boundary is universal and given by the MVP results, Eqs. (\ref{phi_x>0}) and (\ref{1d_density}), that turns nearly-universal, Eqs.(\ref{0_charge_solution})-(\ref{W_density_strong_screening}), at larger distances.  

We argued previously (Section IVA) that the complete screening effect is an artifact - the observable charge of an overcritical source region must be always equal to unity ($1/\alpha$).  Substituting this value into Eq. (\ref{0_charge_solution}) provides us with a length scale
\begin{equation}
\label{W_cloud_size}
R_{W}\simeq a e^{3\pi/2g\alpha}
\end{equation}
which is the radius of the space charge region:  for $r>R_{W}$ the electron density is negligible and the potential is that of unit ($1/\alpha$) charge.  The TF results (\ref{0_charge_solution})-(\ref{W_density_strong_screening}) are applicable at distances $r\ll R_{W}$.  The exact magnitude of the exponential is explained in Section VII.

In QED the exponential factor in (\ref{W_cloud_size}) is about $10^{140}$. Then $R_{W}$ is the largest length scale of the problem and for all practical purposes TF theory is exact in the $Z\gg Z_{c}$ regime.  

In WS with $g=24$ and $\alpha=1/10$ the exponential factor in (\ref{W_cloud_size}) is close to $7$ which means that the whole spatial structure of the overcritical Weyl ion is experimentally accessible.  This system is particularly interesting because both the TF, $r\ll R_{W}$, and the non-TF, $r\gg R_{W}$, regions can be probed.   On the other hand, choosing $g=2$ gives the exponential factor of the order $10^{10}$ which for a nanometer scale impurity region corresponds to the Weyl ion of $10m$ radius.  In the latter case the TF theory provides practically exact description.  

\subsubsection{Narrow-band gap semiconductors and QED}

We already learned that at a distance of a few source radii the potential drops to a value of the order $e\gamma^{-1/2}/\epsilon a$.  This corresponds to the energy scale $e\varphi \simeq \Delta (Z_{xx}/Z)^{1/3}$ which is much larger than the energy gap $\Delta$ if the charge $Z$ is significantly smaller than the characteristic charge
\begin{equation}
\label{xxover_charge}
Z_{xx}\simeq \gamma^{-3/2}\simeq Z_{x}^{3}
\end{equation}
Then the analysis just given for WS will also be applicable in the NBGS/QED case with the conclusion that within several impurity radii the solution to the problem continues to be given by the MVP results, Eqs.(\ref{phi_x<0})-(\ref{1d_density}), with $x\rightarrow r-a$.  

The characteristic charge (\ref{xxover_charge}) separates the regime of moderately strong screening $\gamma^{-1/2}\ll Z \ll \gamma^{-3/2}$ to which the results of this subsection apply, from that of super-strong screening $Z\gg \gamma^{-3/2}$.  The characteristic charge $Z_{xx}$ significantly exceeds the crossover charge $Z_{x}$, Eq.(\ref{weak_strong_xover_charge}), separating the regimes of weak and strong screening response.    In NBGS with $\gamma = 10^{-3}$ we find $Z_{xx}\simeq 30^{3}$ which even by condensed matter standards is very large.  In QED we obtain $Z_{xx}\simeq 3000^{3}$.  In view of unrealistically large value of $Z_{xx}$ the analysis of the regime of super-strong screening $Z\gg Z_{xx}\simeq \gamma^{-3/2}$ is not pursued here.  

At distances exceeding several source radii we need to look at the equation
\begin{equation}
\label{qed_nbgs_Eq_charge}
\frac{dq}{d\ell}=-\frac{2g\alpha}{3\pi}\left ( q^{2}-\frac{2q r}{\Lambda}\right )^{3/2},~~~\ell=\ln\frac{r}{a}
\end{equation}
which generalizes Eq.(\ref{GL}).   Now the charge decreases faster with position than its WS counterpart;  when the right-hand side vanishes, i. e. the edge of the electron shell $r=R_{sc}$  is reached, the charge acquires its observable value $q_{\infty}$ (see Eq.(\ref{qed_observable_charge_definition})) and stops changing thereafter.  The form of the solution can be approximately captured by the WS result (\ref{0_charge_solution}) (that remains relevant at distances $r\ll R_{sc}$);  the value of $q_{\infty}$ at $R_{sc}$ can be estimated with logarithmic accuracy by equating $e \varphi$ to $\Delta$.  This is equivalent to terminating the flow equation (\ref{GL}) at the scale
\begin{equation}
\label{termination_point}
\ell_{sc}=\ln\frac{R_{sc}}{a}\gg1
\end{equation}
and identifying $q({\ell_{sc}})=q_{\infty}$.  Then the observable charge $q_{\infty}$ will satisfy the equation
\begin{equation}
\label{selfcons_obs_charge_eq}
q_{\infty}^{2}\approx \frac{3\pi}{4g\alpha\ln(q_{\infty}\Lambda/a)}
\end{equation}
whose consequence is that for $a$ fixed there exists a nearly universal lower limit on the observable charge $q_{\infty}$.  In the point source limit $a\rightarrow 0$ we find $q_{\infty}=0$, i.e. there is a complete screening of the field of a point charge at an arbitrary distance from it.   For a realistic extended source the approximate solution for the charge is 
\begin{equation}
\label{obs_charge_strong_screening}
q_{\infty} =Q_{\infty}\alpha\approx \sqrt{\frac{3\pi}{4g\alpha \ln(\Lambda/a(g\alpha)^{1/2})}}\rightarrow \sqrt{\frac{9\pi}{4g\alpha \ln(Z_{xx}/Z)}}
\end{equation}
where in the last step we specified to the practically important case of $a\propto Z^{1/3}$.  The size of the space charge region is then given by Eq.(\ref{qed_observable_charge_definition}).  The electric field at the edge of the space charge region can be estimated as 
\begin{equation}
\label{el_field_shell _edge}
E_{sc}=\frac{eq_{\infty}}{\epsilon \alpha R_{sc}^{2}}\simeq\alpha^{1/2}E_{Z}
\end{equation}
The fact that the field at the shell edge (\ref{el_field_shell _edge}) is much smaller than the Zener field demonstrates the sharpness of the edge.

The solution (\ref{obs_charge_strong_screening}) is accurate provided $\ln(Z_{xx}/Z)^{1/3}\gg1$.  For NBGS with $Z\simeq 400$ ($10nm$ impurity region) and $\gamma =10^{-3}$ we find $\ln (Z_{xx}/Z)^{1/3}\approx1.5$.  This is not really in the regime where Eq.(\ref{obs_charge_strong_screening}) applies, but suffices to estimate the charge (ignoring the logarithmic factor) as $Q_{\infty}=q_{\infty}/\alpha \simeq\lambda^{-1/2}\approx 30$, and the size of the space charge region $R_{sc}\simeq 30nm$.  In QED we find $Q_{\infty}\simeq 3000$.   

As the bare charge $Z$ continues to increase within the $\gamma^{-1/2}\ll Z \ll \gamma^{-3/2}$ range of moderately strong screening, the observable charge (\ref{obs_charge_strong_screening}) and size of the space charge region $R_{sc}$ remain nearly constant increasing very slowly with $Z$.  The size of the source region $a\propto Z^{1/3}$ grows faster with $Z$ than $R_{sc}$, at $Z\simeq Z_{xx}$ the two meet, and the result (\ref{obs_charge_strong_screening}) ceases to be applicable.  

To summarize, in NBGS and QED in the regime of moderately strong screening $Z_{x}\ll Z \ll Z_{xx}$ the expressions for the observable charge (\ref{obs_charge_strong_screening}) and radius of the space charge region (\ref{qed_observable_charge_definition}) are nearly-universal.  They are manifestations of the nearly-universal "zero charge" behavior (\ref{0_charge_solution}) in the WS case.  The physical mechanism by which the zero charge situation is avoided is purely classical: when it is no longer energetically favorable, the creation of further space charge terminates.

\subsubsection{Numerical solution}

To put our analysis of the regime of moderately strong screening $\gamma^{-1/2}\ll Z \ll \gamma^{-3/2}$ onto solid footing we solved the full Eq.(\ref{diff_eq_phi}) numerically.  The results are shown in the Figure, where we additionally displayed the charge $Q(r)$ within a sphere of radius $r$ as an indicator of the strength of screening.  
\begin{figure}
\includegraphics[width=1.0\columnwidth, keepaspectratio]{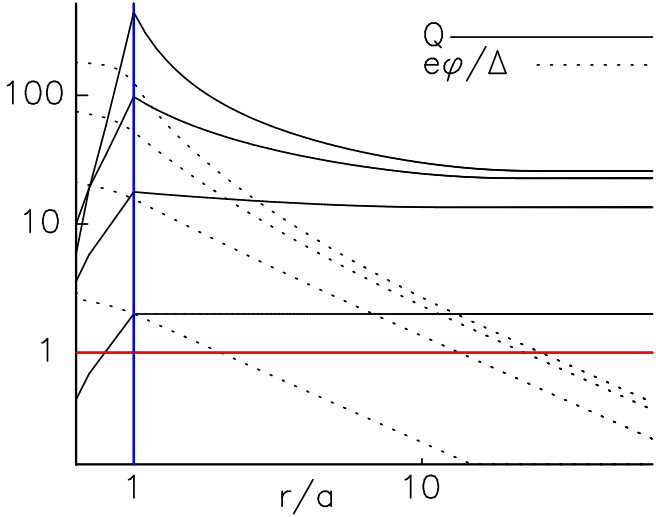} 
\caption{(Color online) Potential $\varphi$ (\ref{definition_potential}) and charge $Q(r)$ within a sphere of radius $r$ (\ref{charge_connection}) as functions of distance (double logarithmic representation).  The source region is $r/a < 1$.  The gap value $\Delta$ is indicated by the horizontal line; the electron cloud is limited to the region where $e\varphi >  \Delta$, which defines $R_{sc}$.  The curves are drawn for $Z = 2, 20, 200, 2000$, $\gamma = 0.001$.  For large $Z$, $R_{sc}$ approaches a $Z$-independent limit, indicating that $Q(r)$ tends to an upper bound $Q_{\infty}$.}
\end{figure} 

The very weak dependence of the observable charge $Q_{\infty}=q_{\infty}/\alpha$ and size $R_{sc}$ of the space charge region on the bare charge $Z$ has its origin in the $Z$-dependence of the size of the source region (\ref{sc_matter}).  For $a=const$, our theory predicts $Z$-independent limit on $Q_{\infty}$ and $R_{sc}$.  Therefore in order to single out this effect we chose $a=const$.  Specifically we set $Z_{0}=\epsilon \Delta a/e^{2}=1$ (other values of $Z_{0}$ are equivalent to a rescaling $Z\rightarrow Z/Z_{0}$ and $\gamma\rightarrow\gamma Z_{0}^{2}$).  Several features of the numerical solution illustrating our analysis deserve mentioning:

(i)  As expected the screening effect of the space charge becomes noticeable for $\gamma Z^{2}\gtrsim1$.

(ii)  The crossing of the charge curves for $Z = 200$ and $Z = 2000$ at $r$ small  is a direct illustration of screening: the TF screening length $\kappa^{-1}$ (see Eq.(\ref{TF_length}) for $a=const$) is smaller for $Z = 2000$ than $Z = 200$, so that the screening at the central region is more complete in the former case.

(iii)  The large drop of the potential within a few radii of the impurity, clearly seen in the $Z=200$ and $Z=2000$ curves, is an illustration of our observation made in the analysis of the WS problem that within this range the potential must drop by a large factor of $(\gamma Z^{2})^{1/6}$.  

(iv)  Remarkably, for $Z\gg1$ there exists a $Z$-independent limit on $R_{sc}$ and $Q_{\infty}$ whose values agree with our estimates.

\section{Weak screening regime $\lambda Z^{2}\ll1$ and synthesis }

The analysis carried out so far relied on the concept of the TF screening length $\kappa^{-1}$, Eq.(\ref{TF_length}), which in the weak-screening regime $\gamma Z^{2}\ll1$ loses its meaning as a length scale characterizing the source region, and one has to start anew.  On the other hand, weak screening means that electron-electron interactions can (possibly) be  treated by a perturbation theory.  To lowest order in $\gamma \ll 1$ we set  $\varphi =\varphi_{ext}$, and then Eq.(\ref{n_of_phi}) gives
\begin{equation}
\label{density_weak screening}
n(\textbf{r})= \frac{\gamma}{4\pi} \Bigg\{\frac{\epsilon \varphi_{ext}(\textbf{r})}{e}\left (\frac{\epsilon \varphi_{ext}(\textbf{r})}{e}-\frac{2}{R_{e}}\right )\Bigg\}^{3/2}
\end{equation}
For $r\leqslant a$ we have $\epsilon \varphi_{ext}/e\simeq Z/a\simeq Z^{2/3}/R_{e}\gg 1/R_{e}$.  Then the density of the space charge inside the source region can be estimated as $n\simeq \gamma Z^{3}/a^{3}$, implying that the number of electrons residing at $r\leqslant a$, is of the order $\gamma Z^{3}$.  The latter must be much smaller than the bare charge $Z$ (to justify the approximation $\varphi=\varphi_{ext}$) thus specifying the condition of weak screening as $\gamma Z^{2}\ll1$.

Outside of the impurity region Eq.(\ref{density_weak screening}) becomes 
\begin{equation}
\label{outside_n_weak screening}
n(r)=\frac{\gamma Z^{3}}{4\pi r^{3}}\left (1-\frac{2r}{z\Lambda}\right )^{3/2}, ~r \leqslant R_{sc}=\frac{z\Lambda}{2}
\end{equation}
and $n=0$ otherwise.  The total screening charge is then of order $\gamma Z^3 \ll Z$, so that $q_{\infty}=z$ (or $Q_{\infty}=Z$), consistent with the TF relationship, Eq.(\ref{qed_observable_charge_definition}).  The electric field at the boundary of the space charge region can be estimated as 
\begin{equation}
\label{el_field_weak_sreening}
E(R_{sc})=\frac{q_{\infty}e}{\epsilon \alpha R_{sc}^{2}}\simeq \frac{1}{z}E_{Z}
\end{equation}
which in view of the condition $z\gg1$ demonstrates the sharpness of the boundary of the space charge region. 

Since the space charge residing at $r\leqslant a$ is small, Eq.(\ref{outside_n_weak screening}) can be used to compute with logarithmic accuracy the net charge $q(r)$ within a sphere of radius $r>a$.

\subsection{Weyl semimetal}

In the WS case when $R_{e}=\infty$ the density of the space charge is given by
\begin{equation}
\label{W_density_small distances}
n(r)=\frac{\gamma Z^{3}}{4\pi r^{3}}=\frac{gz^{3}}{6\pi^{2}r^{3}}
\end{equation}
and we find 
\begin{equation}
\label{charge_weak_screening}
q(r)\approx z-4\pi \alpha\int_{a}^{r}y^{2}n(y)dy= z\left (1-\frac{2g\alpha z^{2}}{3\pi} \ln\frac{r}{a}\right )
\end{equation}
This expression is applicable provided $(2g\alpha z^{2}/3\pi) \ln(r/a) \ll 1$, i.e. it inevitably fails at sufficiently large distance from the source.

Alternatively, the weak screening $\gamma Z^{2}\ll1$ analysis can be carried out by treating the cubic term of (\ref{diff_eq_chi_of_x}) perturbatively.  Then the lowest order solution outside the source is $\chi \alpha=z$.  The next order gives for $r>a$
\begin{equation}
\label{chi_weak screening}
\chi \alpha=z\left (1-\frac{2g\alpha z^{2}}{3\pi} \ell \right )=z\left (1-\frac{2g\alpha z^{2}}{3\pi} \ln\frac{r}{a} \right )
\end{equation}
We observe that the expression for charge (\ref{charge_connection}) (in natural units) computed with the help of Eq.(\ref{chi_weak screening}) agrees with Eq.(\ref{charge_weak_screening}) to logarithmic accuracy which we adopt.  Then the perturbative expression (\ref{chi_weak screening}) may be regarded as a charge itself:  it tells us that within the cloud, the physical potential $\varphi$ and the density of space charge $n$ decrease with $r$ faster than $1/r$ and $1/r^{3}$, respectively.  

On the other hand, no matter what the strength of screening is, at sufficiently large distances from the source center the charge $q(r)$ is given by the asymptotic limit of Eq.(\ref{0_charge_solution}).  All these results can be summarized in a simple interpolation formula for the charge $q=\chi \alpha$
\begin{equation}
\label{interpolation_charge}
q^{2}(r)=\frac{z^{2}}{1+ (4g\alpha z^{2}/3\pi)\ln(r/a)}
\end{equation}
If the parameter $z$ is viewed more broadly as the \textit{net} charge within the source region, then this equation with $z=z^{*}\simeq \alpha^{-1/2}$ also covers the regime of strong screening $\gamma Z^{2}\gg 1$ (see Eq.(\ref{0_charge_solution})).  This choice additionally guarantees that the results of Section VB2 pertinent to the NBGS/QED case are automatically captured.

Substituting $\chi=q/\alpha$ into Eqs.(\ref{def_chi}) and (\ref{n_of_phi}) we find corresponding interpolation formulas for the potential
\begin{equation}
\label{interpolation_potential}
\varphi(r)=\frac {Ze}{\epsilon r\sqrt {1+ (4g\alpha z^{2}/3\pi)\ln(r/a)}}
\end{equation}
and the electron density
\begin{equation}
\label{interpolation_density}
n(r)=\frac{gz^{3}}{6\pi^{2} r^{3}[1+(4g\alpha z^{2}/3\pi)\ln(r/a)]^{3/2}}
\end{equation}
The logarithmic terms in the denominators of Eqs.(\ref{interpolation_charge})-(\ref{interpolation_density}) are relevant at the scales $r$ exceeding
\begin{equation}
\label{shell_screening_length}
R_{scr}\simeq a e^{3\pi/4g\alpha z^{2}}
\end{equation}
This quantity is the screening length within the space charge of Weyl electrons. Deviations from the Coulomb law become substantial for $r>R_{scr}$ and the asymptotic regimes of Eqs.(\ref{0_charge_solution})-(\ref{W_density_strong_screening}) are reached at $r\gg R_{scr}$.  Specifically, as the strength of screening increases from small to large $\gamma Z^{2}$ the screening radius (\ref{shell_screening_length})  decreases from a very large value to the scale comparable to the source size.

Since the zero charge effect is an artifact, consistency of the theory requires that the screening length (\ref{shell_screening_length}) to be significantly shorter than the radius of the electron cloud (\ref{W_cloud_size}).  Since $z\gg1$ this is indeed true.

\subsection{Narrow-band gap semiconductors and QED}

The NBGS/QED case will be handled in exactly the same manner as that of the regime of moderately strong screening - by terminating the WS solution at the scale of the space charge region (\ref{termination_point}).  Then the observable charge follows from Eq.(\ref{interpolation_charge}) as 
\begin{equation}
\label{sc_interpolation_charge}
q_{\infty}^{2}\approx\frac{z^{2}}{1+ (4g\alpha z^{2}/3\pi) \ln(q_{\infty}\Lambda/a)}
\end{equation}
We see that the initial growth of $q_{\infty}(z)$ as $z$, in the regime of weak screening $\gamma Z^{2}\simeq \alpha z^{2}\ll1$, slows down eventually saturating, in the strong screening regime $\gamma Z^{2}\simeq \alpha z^{2}\gg1$, at $z$-independent value implied by Eq.(\ref{selfcons_obs_charge_eq}).  In the regime of weak screening the solution to Eq.(\ref{sc_interpolation_charge}) one step beyond the zero order $q_{\infty}\approx z[1-(2g\alpha z^{2}/3\pi)\ln(z\Lambda/a)]$ reproduces previous findings \cite{MVP}.  

The dependence of the observable charge $Q_{\infty}$ of a supercharged heavy nucleus on the bare charge $Z$ was evaluated in Ref. \cite{numerical} by numerically solving the TF theory discussed in our paper.  The     $Q_{\infty}(Z)$-dependence was found to be a monotonically increasing function with growth rate decreasing with $Z$;  for $Z\rightarrow \infty$ the function $Q_{\infty}(Z)$ was found to grow slower than $Z$.   The MVP theory \cite{MVP} explained the $Q_{\infty}(Z)$ behavior in the regime of weak screening $\gamma Z^{2}\ll 1$ and laid out a foundation to understand the regime of strong screening $\gamma Z^{2}\gg1$;  its place in the problem of screening of overcritical external charge was explained earlier.  However the zero-charge type solution of the TF theory in the WS case was missed whose consequences are:  

(i) for $a$ fixed the observable charge $Q_{\infty}$ saturates as $Z\rightarrow \infty$ at a $Z$-independent  value;

(ii)  for realistic $a\propto Z^{1/3}$ the $Q_{\infty}(Z)$ dependence is a nearly universal slowly increasing function of $Z$, Eq.(\ref{obs_charge_strong_screening}), which goes beyond explanation of numerical results \cite{numerical}.

\section{Deficiencies and improvement of the Thomas-Fermi theory}

Even though in the NBGS/QED case complete screening does not occur, there are other respects in which the TF theory is internally inconsistent.  Solving Eq.(\ref{sc_interpolation_charge}) for the bare charge $z$ we find
\begin{equation}
\label{Landau_pole}
z^{2}\approx\frac{q_{\infty}^{2}}{1- (4g\alpha q_{\infty}^{2}/3\pi) \ln(q_{\infty}\Lambda/a)}
\end{equation}  
While correctly predicting that the latter is always larger than the observable charge, Eq.(\ref{Landau_pole}) also tells us that for fixed $q_{\infty}$ and $a\rightarrow 0$ the denominator vanishes for finite $a$ given by 
\begin{equation}
\label{pole_position}
a_{p}\simeq \Lambda q_{\infty} e^{-3\pi/4g\alpha q_{\infty}^{2}}
\end{equation}
before the limit of point source is reached. At $a=a_{p}$ the bare charge is infinite while for $a<a_{p}$ it is imaginary; the latter feature is certainly unacceptable.  These conclusions having their origin in the "zero charge" solution in the WS case, like the zero charge effect itself, are artifacts.  Even if the vanishing of the observable charge did occur, the scale (\ref{pole_position}) is too small to be of any practical importance.  

In Section IVA we introduced a modified TF equation (\ref{modified_TF}) with built-in quantum-mechanical fall to the center.  It is straightforward to realize that such a modification removes the zero charge effect in the WS case;  the NBGS/QED problems are also liberated of the difficulty of the vanishing denominator.  Specifically, in the WS ($\Delta=0$) case substituting (\ref{def_chi}) into (\ref{modified_TF})  we obtain (instead of (\ref{GL})) the following flow equation
\begin{equation}
\label{W_flow_eq}
\frac{dq}{d\ell}=-\frac{2g\alpha}{3\pi}(q^{2}-1)^{3/2}
\end{equation}
Now any "initial" charge $q(0)=z>1$ will be carried to the stable fixed point $q=1$ which is reached as $\ell \rightarrow \infty$.  Eq.(\ref{W_flow_eq}) can be integrated in quadratures for arbitrary initial $z$ with the result
\begin{equation}
\label{better_W_solution}
q^{2}(\ell)-1=\frac{1}{(2g\alpha/3\pi)^{2}(\ell+\ell_{W})^{2}-1}
\end{equation}
where the scale $\ell_{W}=3\pi z/(2g\alpha\sqrt{z^{2}-1})$ is determined by the condition $q(\ell=0) = z$. The corresponding spatial scale
\begin{equation}
\label{better_W_cloud_size}
R_{W}= a e^{\ell_{W}}=a \exp\left (\frac{3\pi}{2g\alpha}\frac{z}{\sqrt{z^{2}-1}}\right )
\end{equation}
characterizes the size of the space charge cloud.  For $r\gg R_{W}$ the (overcritical) impurity charge appears as being poised at the critical value.  As $z\rightarrow 1+0$ (a weakly overcritical source), the cloud size (\ref{better_W_cloud_size}) diverges because the creation of spacial charge is a critical phenomenon.  The Kosterlitz-Thouless-type essential singularity in (\ref{better_W_cloud_size}) is typical of localization transitions related to the quantum-mechanical fall to the center \cite{KS}.  For the supercritical source $z\gg1$ the cloud size becomes Eq.(\ref{W_cloud_size}).

The point source limit $a\rightarrow 0$ ($\ell=\infty$) of the solution (\ref{better_W_solution}) also describes the NBGS/QED problems with the conclusion that at arbitrary distance $r$ from overcritical point source the observable charge of the latter is unity.    

\section{ACKNOWLEDGMENT}

This work was supported by US AFOSR Grant No. FA9550-11-1-0297.

\end{document}